\documentclass{elsart}
\usepackage{amssymb}
\usepackage{graphicx}

\journal{Nucl.Instrum.Meth. Section A}
\begin{document}
\begin{frontmatter}

\title{Double-Layer Silicon PIN Photodiode X-Ray Detector for a Future
X-ray Timing Mission}
\author[a]{Hua Feng\corauthref{cor}},
\corauth[cor]{Corresponding author.}
\ead{hua-feng@uiowa.edu}
\author[a]{Philip Kaaret},
\author[b]{Hans Andersson}
\address[a]{Department of Physics and Astronomy, University of Iowa, Van
Allen Hall, Iowa City, IA 52242}
\address[b]{Oxford Instruments Analytical Oy, P.O. Box 85, 02631 Espoo, Finland}

\begin{abstract}

A double-layer silicon detector consisting of two 500$\mu$m-thick
silicon PIN photodiodes with independent readouts was mounted in a
vacuum chamber and tested with X-ray sources.  The detector is
sensitive from 1--30 keV with an effective area of 6 mm$^2$.  The
detector performs best at $-$35 $^\circ$C with an energy resolution of
220~eV (FWHM, full width at half maximum) at 5.9~keV, and is able to
operate at room temperature, $+$25 $^\circ$C, with moderate resolution
around 760~eV (FWHM).  The response of the top layer sensor is highly
uniform across the sensitive area.  This large-format silicon detector
is appropriate for future X-ray timing missions.

\end{abstract}


\end{frontmatter}

\section{Introduction}

New detectors with large effective areas, rapid timing response, and
good sensitivity up to $\sim$30~keV  are required to fulfill the
scientific goals for a future X-ray timing observatory \cite{kaa04}
beyond the Rossi X-ray Timing Explorer (RXTE) \cite{bra93}.  RXTE has
been operating for ten years with fruitful scientific results obtained
from its Proportional Counter Array (PCA), which is close to the end of
its lifetime.  Thick silicon detectors are, perhaps, the best candidate
detectors for future timing missions.  Silicon PIN photodiodes are
simple to fabricate, with mature applications in industry and science \cite{mit04},
and could be manufactured in large volumes at moderate cost with
existing technology.  Large-format thick silicon PIN diode arrays
sensitive in 1-30 keV would give spectral resolutions of a few hundred
eV (FWHM) at 6~keV, which is better than that of PCA.   Also, solid
state detectors should have good long-term stability.

This paper describes the structure and test of a double-layer
silicon PIN detector.  We present measurement of the spectral
resolution and its variation with temperature, and the uniformity of
the detector efficiency, charge collection, and resolution.  We
conclude with a summary of the performance of the detector and the
important characteristics for its application in X-ray astronomy.

\section{Detector and Experiment Configuration}

A double-layer silicon detector was assembled from two square silicon
chips with high voltage applied to both chips on the middle layer
between the chips and guard rings on both chips that create a circular
effective area with a diameter of 2.75~mm.  This silicon detector
sandwich is attached to an alumina substrate along with two FETs acting as
the front stage of the detector preamplifiers.  A peltier cooler
(MI2012T-01AC from Marlow Industries, Inc.) is adhered on the other
side of the alumina substrate using silver epoxy.  The warm side of the
cooler is connected to a copper heat sink which is mounted in the
center of a CF-100 flange.  Thin wires connected to the detector to
pass high voltage and electronic signals and another two, thicker,
wires to power the cooler are connected via feedthroughs in the flange
to an electronics box mounted outside the vacuum regions and containing
two preamplifiers and high voltage filtering circuitry.  

The preamplifier is charge sensitive with a reset switch, instead of a
feedback resistor, to discharge the feedback capacitor.  The two
preamplifier outputs for both layers are connected to shaping
amplifiers with a gain of 1000 and adjustable shaping time.  The shaped
outputs are stretched to a pulse duration of 5~$\mu$s and the stretched
outputs are fed into a Wavebook (Wavebook 512A from IOtech, Inc.) for
pulse digitization.  The shaped pulse from each channel is fed into a
single channel analyzer used to trigger the readout.  A pulse above
threshold from either channel will trigger the Wavebook to sample the 
pulses from both channels.

The CF-100 flange with the detector is mounted on the side of a vacuum
chamber.   An X-ray tube is located on the other side of the vacuum
chamber on the same axis as the detector.  A steel mask with a 0.33 mm
diameter central hole is placed in front of the detector as a
collimator.  It is carried by a two dimensional linear stage in order
to move the exposure point at different positions above the sensitive
area.  Radioactive source of $^{55}$Fe and $^{109}$Cd are mounted on
the same mask and may be positioned in front of the detector as
required.

\section{Spectral Resolution}

\begin{figure}
\centering
\includegraphics[width=0.5\textwidth]{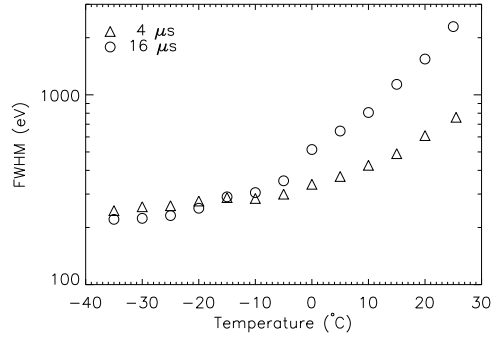}
\caption{Energy resolution (FWHM) vs. temperature for $^{55}$Fe
K$\alpha$ line with the shaping time of 4 $\mu$s and 16 $\mu$s
respectively. 
\label{fig:temp}}
\end{figure}

We obtained $^{55}$Fe spectra at various temperatures in the range
$-$35 to $+$25$^\circ$C with shaping times of 4 $\mu$s and 16 $\mu$s. 
The energy resolution versus temperature is shown in
Fig.~\ref{fig:temp}. At high temperatures, a smaller shaping time such
as 1 $\mu$s gives better resolution, but also cause oscillations in the
shaping amplifier output, which produces much more noise at low
energies. At low temperature around $-$35 $^\circ$C, a shaping time of
16 $\mu$s produces the best resolution, which, though, is not strongly
influenced by changing the shaping time in the range 8--24 $\mu$s.

\begin{figure}
\centering
\includegraphics[width=0.49\textwidth]{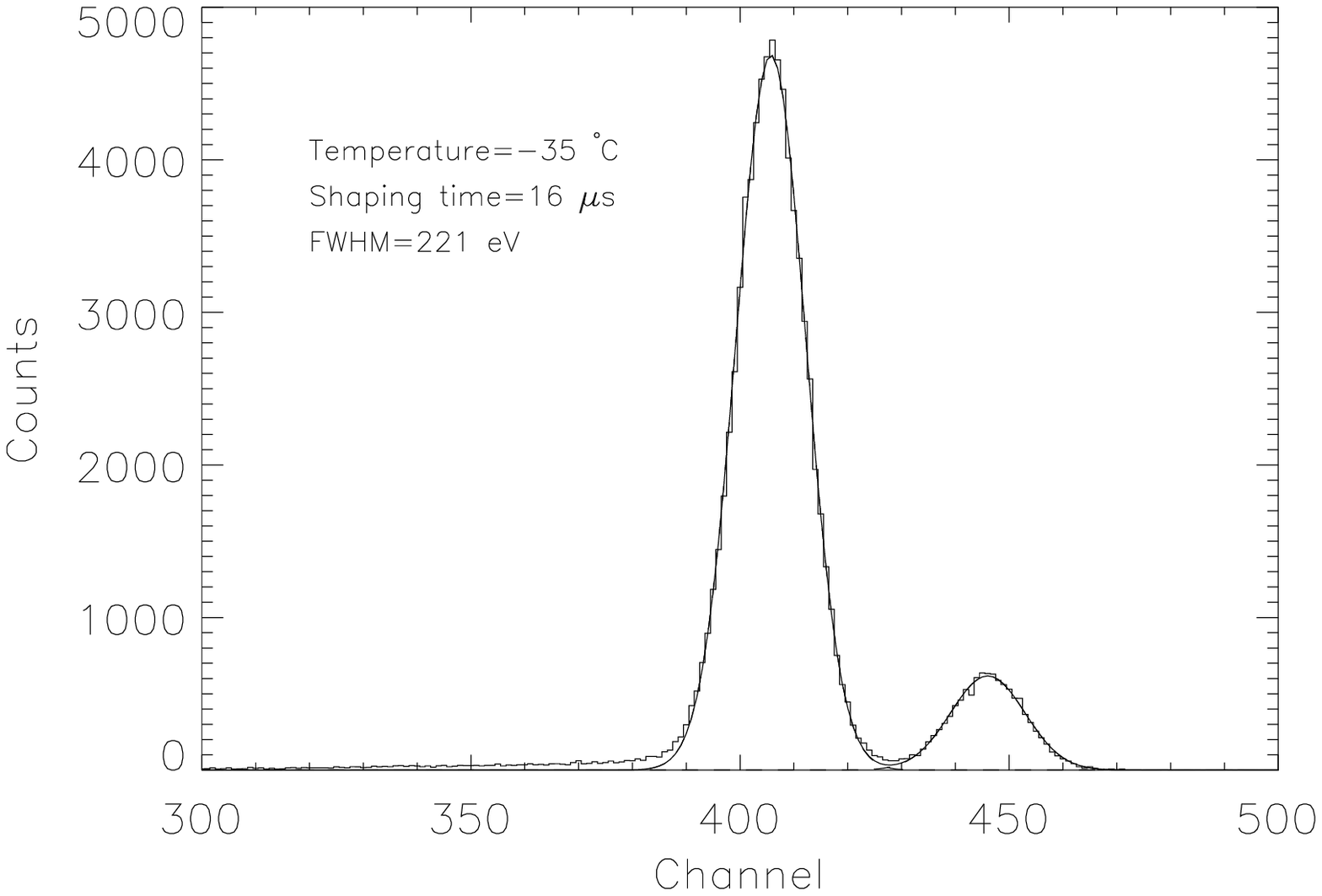}\\
\includegraphics[width=0.49\textwidth]{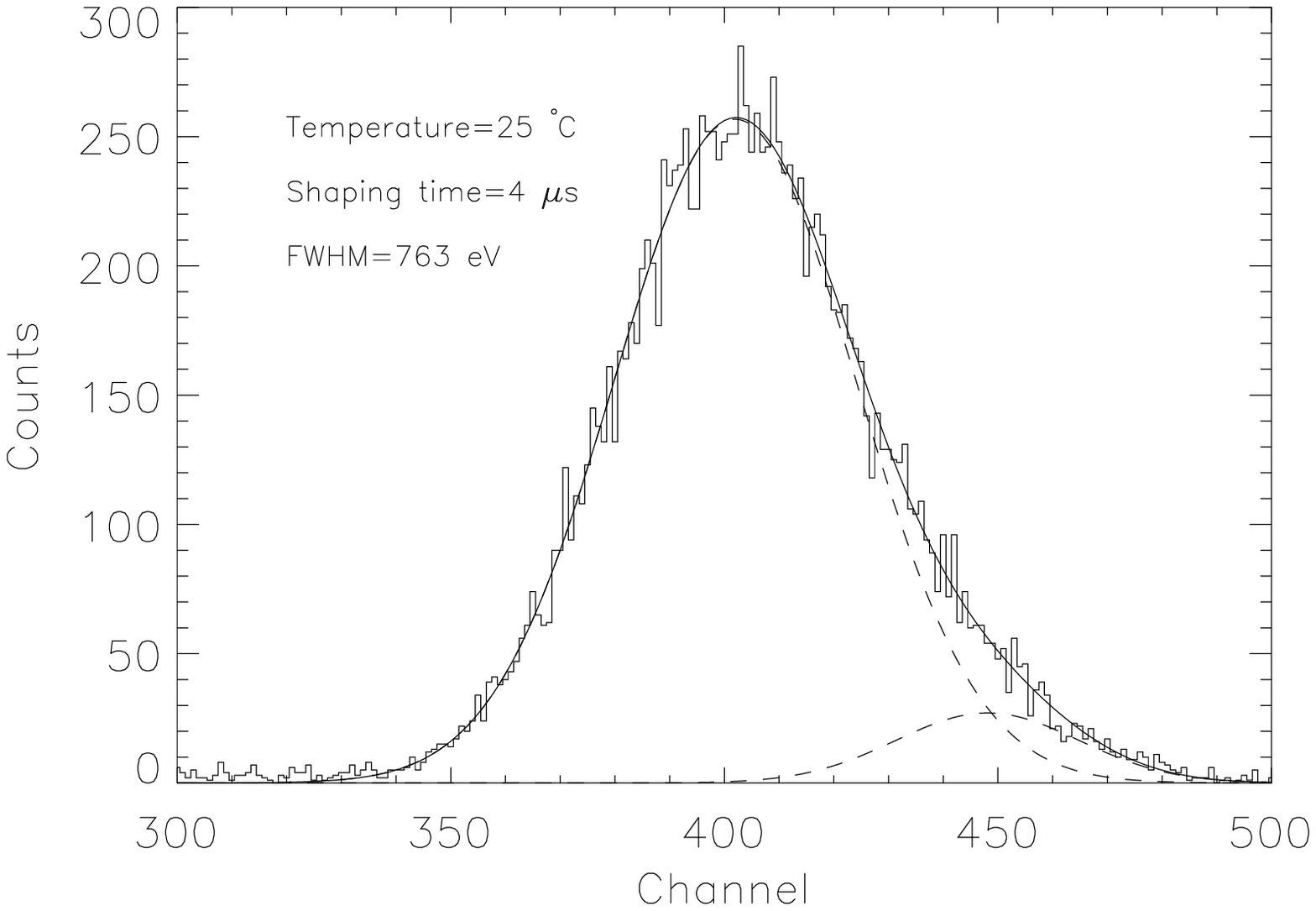}
\caption{Energy spectra of $^{55}$Fe. The best fitted spectrum with two
Gaussian functions is plotted with dashed lines representing the Mn
K$\alpha$ and K$\beta$ lines respectively. Top: best performance at
temperature of $-$35 $^\circ$C and shaping time of 16 $\mu$s. Bottom:
at room temperature with the shaping time of 4 $\mu$s.
\label{fig:specfit}}
\end{figure}

Fig.~\ref{fig:temp} shows that the best resolution is obtained at $-$35
$^\circ$C with the shaping time of 16 $\mu$s. A moderate
resolution of 763 eV is obtained at temperature of 25 $^\circ$C with a
shaping time of 4 $\mu$s. These two spectra are shown in Fig.~\ref{fig:specfit}.
We fitted each  spectrum with a model
consisting of the sum of two Gaussian functions, corresponding to the
Mn K$\alpha$ (5.9 keV) and K$\beta$ (6.5 keV) lines respectively.  
This model provided an adequate fit to all of the spectra.  

\begin{figure}
\centering
\includegraphics[width=0.49\textwidth]{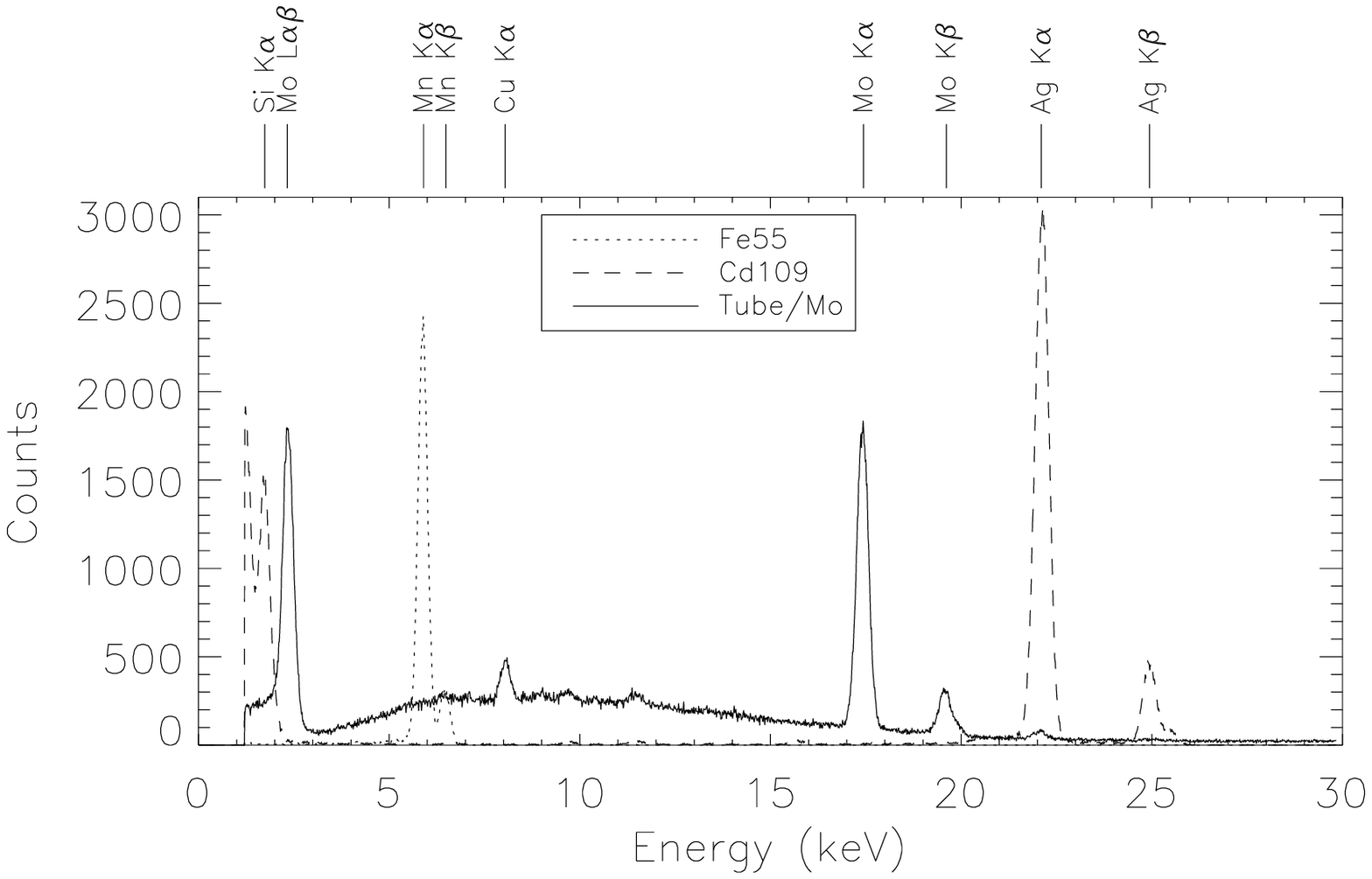}\\
\includegraphics[width=0.49\textwidth]{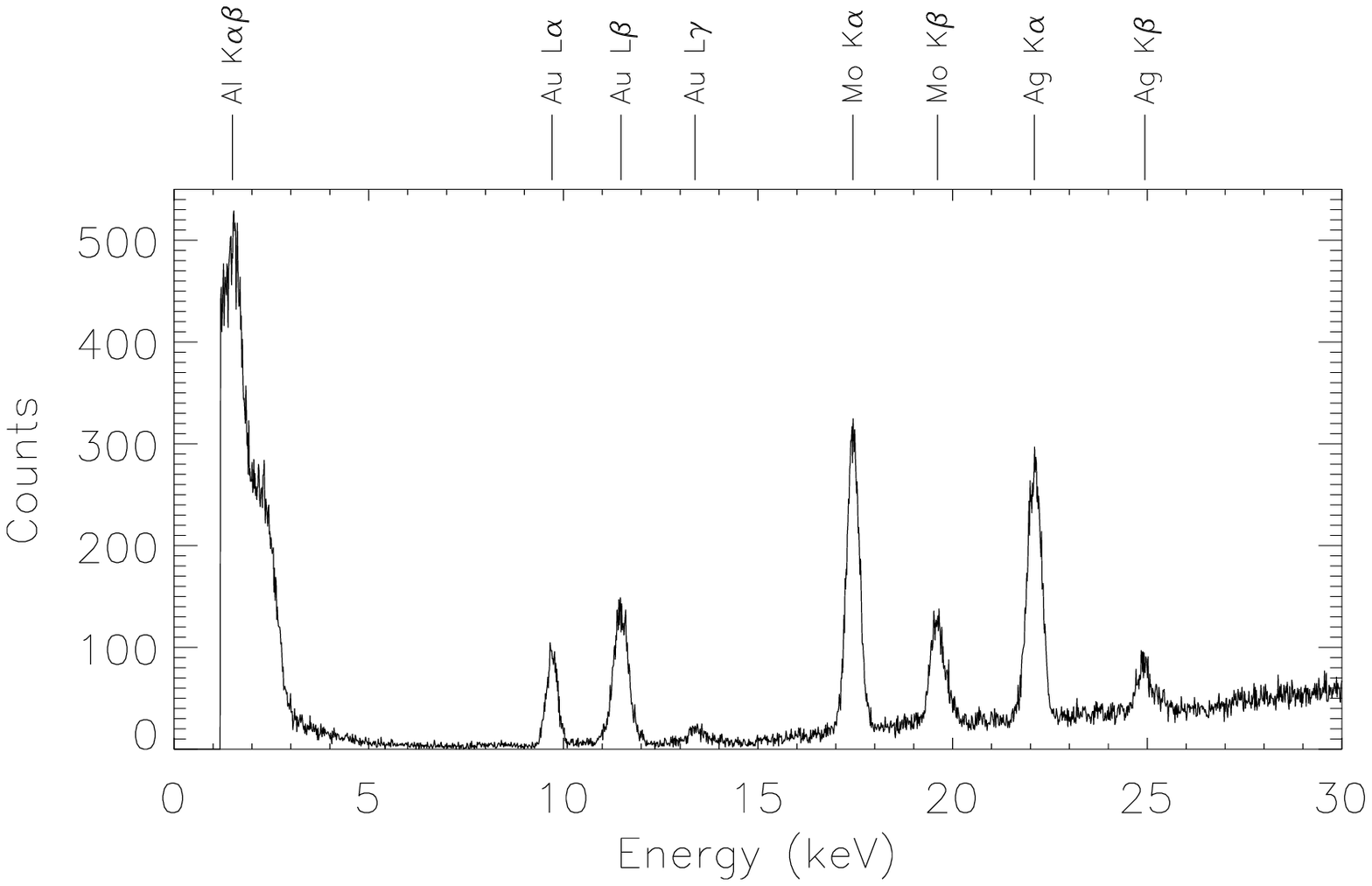}
\caption{Energy Spectra from $^{55}$Fe, $^{109}$Cd and the X-ray tube
with Mo as the target. Emission lines are labeled on the top. Line
energies and widths are listed in Table~1. Top: Top layer
spectra; Bottom: Bottom layer spectrum from the X-ray tube only.
\label{fig:spec}}
\end{figure}

\begin{figure}
\centering
\includegraphics[width=0.49\textwidth]{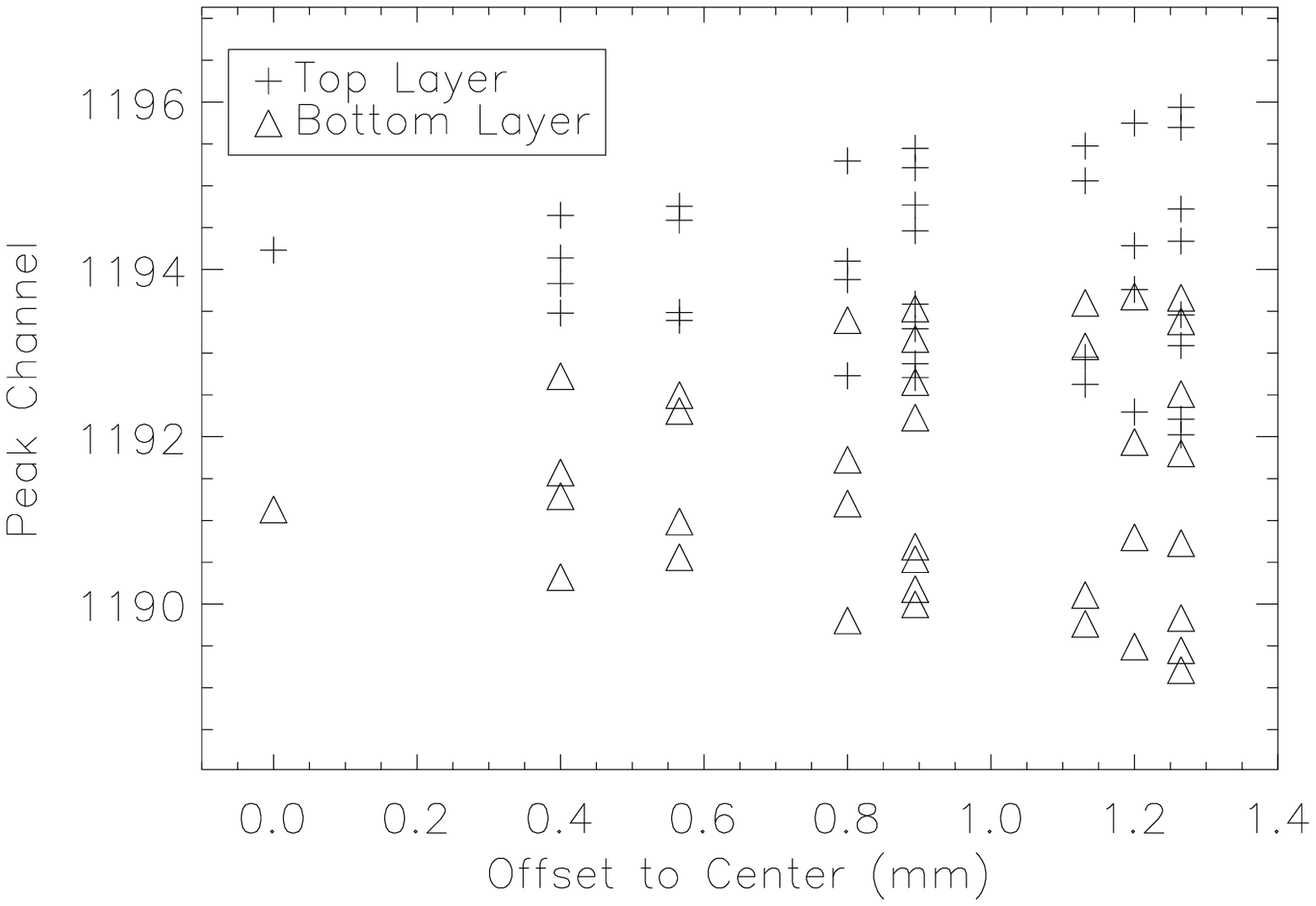}\\
\includegraphics[width=0.49\textwidth]{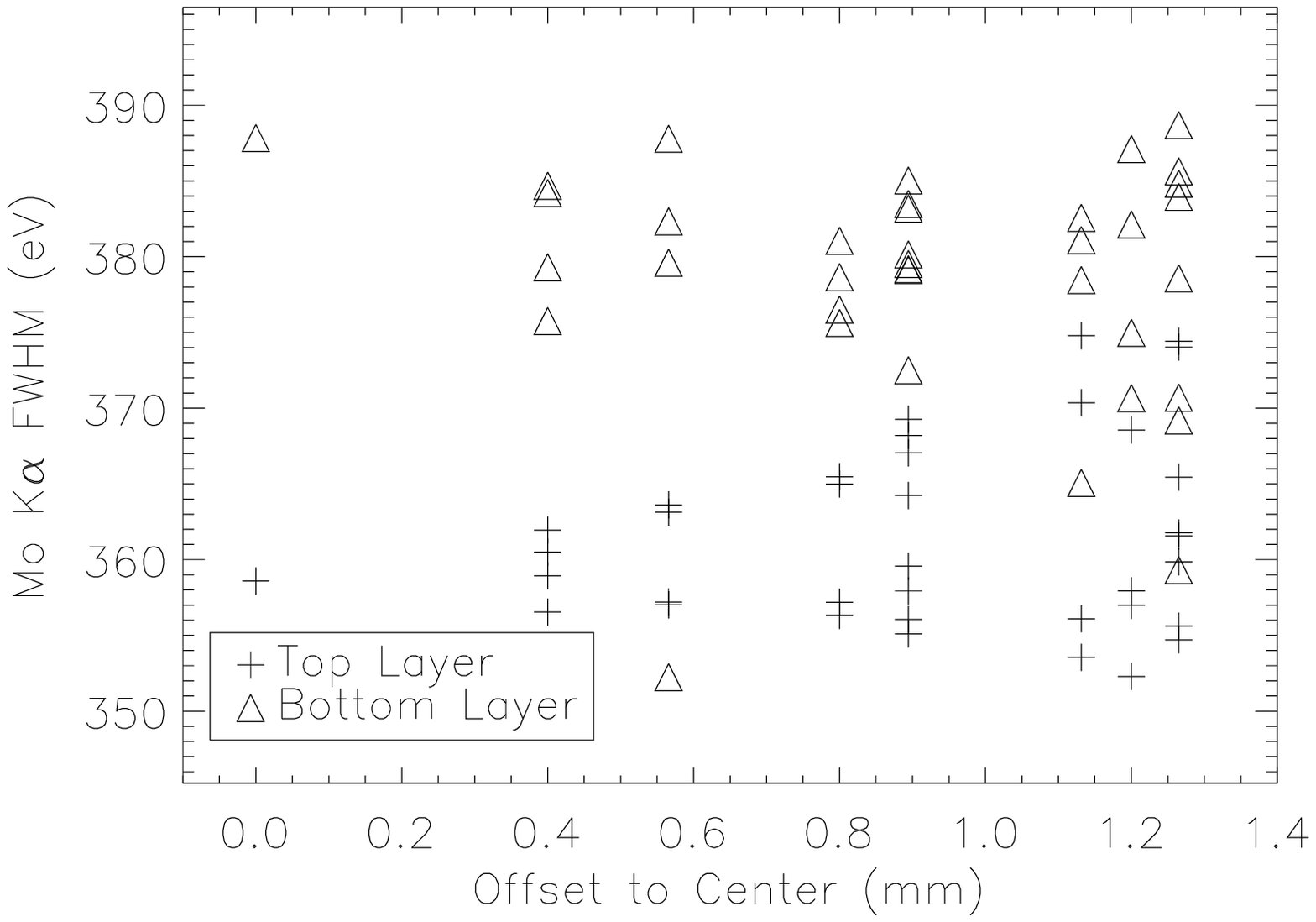}
\caption{Mo K$\alpha$ line peak channel (top panel) and FWHM (bottom
panel) at different spots across the detector for both layers. 
\label{fig:unie}}
\end{figure}

In an early measurement there appeared a notable condensation 
problem, which made the resolution become worse unexpectedly at temperatures 
below $-$20 $^\circ$C. To reduce the condensation, we added 
another cooler inside the chamber and made it below $-$40 $^\circ$C
for several days prior to the measurement. This helped much but would
not prevent condensation if the detector stayed at low temperature
longer than one hour. Therefore, every spectrum below 0 $^\circ$C in Fig.~\ref{fig:temp} \& \ref{fig:specfit} was obtained when the detector
was cooled down directly from room temperature within one hour. While 
spectra from Fig.~\ref{fig:spec} to Fig.~\ref{fig:unie} were obtained
when the temperature stayed at $-$20 $^\circ$C for long time since those
results did not suffer from the condensation problem much.

The electronic noise \cite{iwa95} can be described as

\begin{equation}
\label{equ:noise}
\Delta E(eV)=2.355\frac{w}{q}\left(
c_0qI_L +
c_12kTC_{in}^2/g_m +
c_2A_{1/f} + 
c_3C_{in}^2
\right)^\frac{1}{2},
\end{equation}

\noindent where $w$ is 3.6 eV/e-h for silicon, $q$ is the electron charge, $I_L$ is the leakage current, $k$ is the Boltzmann constant, $T$ is the temperature,
$C_{in}$ is the total input capacitance, $g_m$ is the trasconductance of the input FET, $A_{1/f}$ is the excess $1/f$ noise, $c_{0-3}$ are
coefficients for different noise power terms, $c_0$ is proportional 
to the shaping time and $c_1$ is inversely proportional 
to the shaping time. With a given shaping time,
$c_2A_{1/f}+c_3C_{in}^2$ is a constant and the leakage current can be
expressed as a exponential function. Therefore Eq.~(\ref{equ:noise})
can be re-written as 

\begin{equation}
\label{equ:fit}
\Delta E(eV)=2.355\frac{w}{q}\left(a_0\e^{a_1T}+a_2T+a_3\right)^\frac{1}{2},
\end{equation}

\noindent where $a_{0-3}$ are constants. 

We fit the two curves in Fig.~\ref{fig:temp} corresponding to the two
different shaping times to Eq.~(\ref{equ:fit}) and obtained two sets of
coefficients $a_{0-3}$.  We find that the ratio $a_{0,16\mu {\rm
s}}/a_{0,4\mu {\rm s}} = 4.4$, which is close to the ratio of the two
shaping times as expected.  The coefficient $a_1$ is primarily
determined by the temperature dependence of the leakage current.  For a
shaping time of $4\mu$s, we find $a_{1,4\mu {\rm s}}=0.11$
corresponding to an increase in leakage current by a factor of 1.1 for
an increase in temperature of 1~$^\circ$C.  This is consistent with our
direct measurement of the temperature dependence of the leakage current
and is close to that expected for silicon and measured for other Si PIN
photodiodes.  For a shaping time of $16\mu$s, we find $a_{1,16\mu {\rm
s}}=0.14$ which indicates a somewhat faster increase in the noise
versus temperature.  The determination of the linear component, i.e.
$a_2$, is  of high uncertainty, because the resolution tends to be a
constant at  low temperatures. The fitting results do not change
significantly whether  we force $a_2=0$ or not.

Applying a temperature of $-$20~$^\circ$C and a shaping time of
$16\mu$s, we obtained spectra using the radioactive sources $^{55}$Fe
and $^{109}$Cd and using an X-ray tube with Mo target.  The $^{55}$Fe
radioactive source shows Mn K$\alpha$ and K$\beta$ lines only on the
top layer.  The $^{109}$Cd radioactive source presents Ag K$\alpha$ and
K$\beta$ lines on both layers and a Si K$\alpha$ line on the top
layer.  The latter is caused by X-rays striking the Si outside the
sensitive region and producing a fluorescence X-ray which then
interacts within the sensitive region. The X-ray tube was operated at
voltage of 40~kV  and produces bremsstrahlung radiation and the
characteristic emission lines from the Mo target and a Cu tube holding
the window.  In the bottom layer spectrum, there are strong Au lines
from fluorescence in a Au strip located between the two detector
layers.  This line can also be seen in the top layer spectrum, but is
not as prominent as in the bottom layer due to the high flux from the
bremsstrahlung spectrum.  The Ag lines in the tube spectrum arise from
the silver epoxy centered and beneath the sandwich.  Al lines from the
bottom layer spectrum are X-rays reprocessing on the alumina substrate
beneath the detector.  The measured energy spectra are shown in
Fig.~\ref{fig:spec} for the top layer with three sources and for the
bottom layer with the X-ray tube only. There are 13 lines detected
which are labeled on Fig.~\ref{fig:spec} and their energies and widths
are listed in Table~1.

\section{Uniformity}

\begin{figure}
\centering
\includegraphics[width=0.49\textwidth]{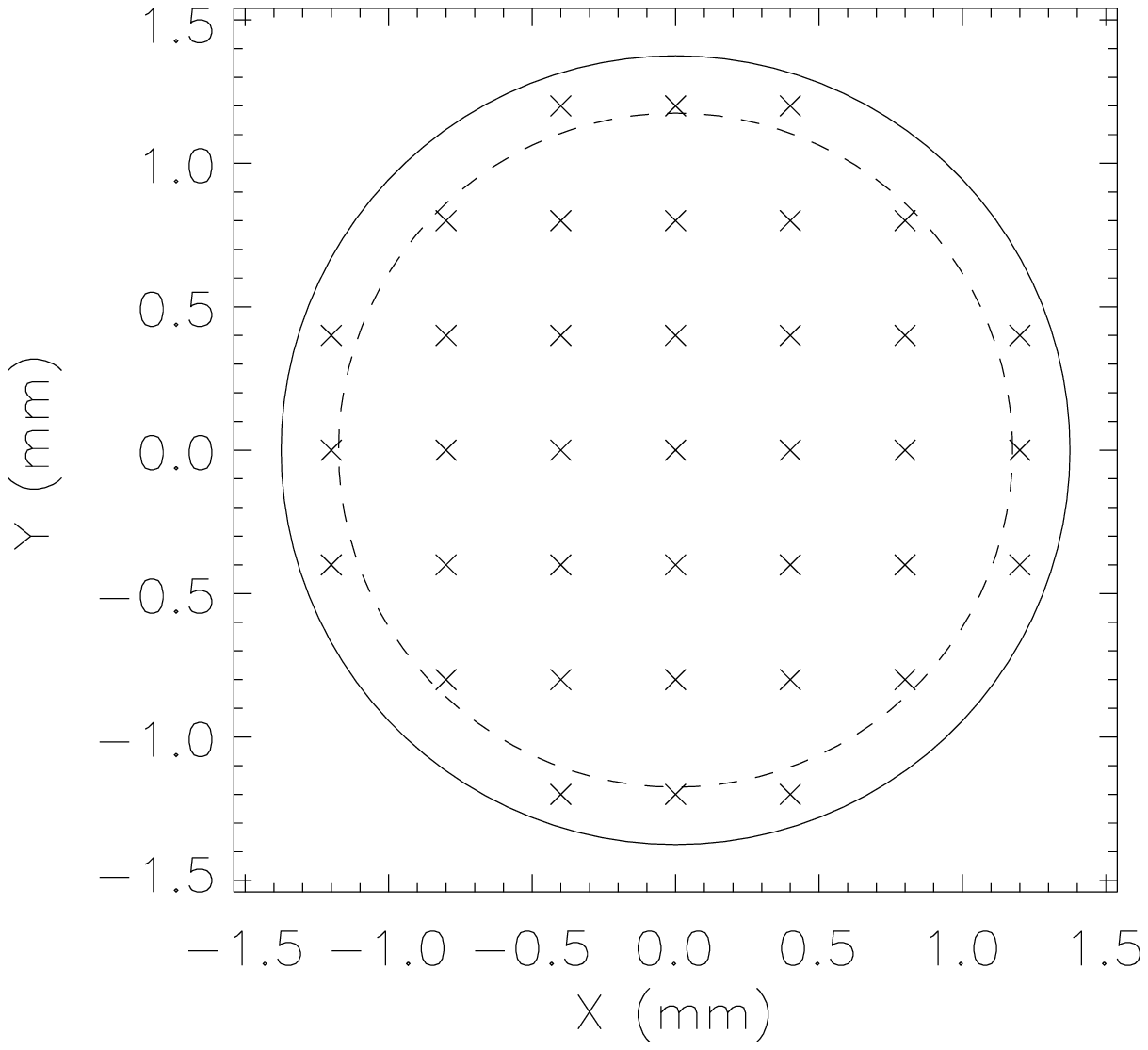}
\includegraphics[width=0.49\textwidth]{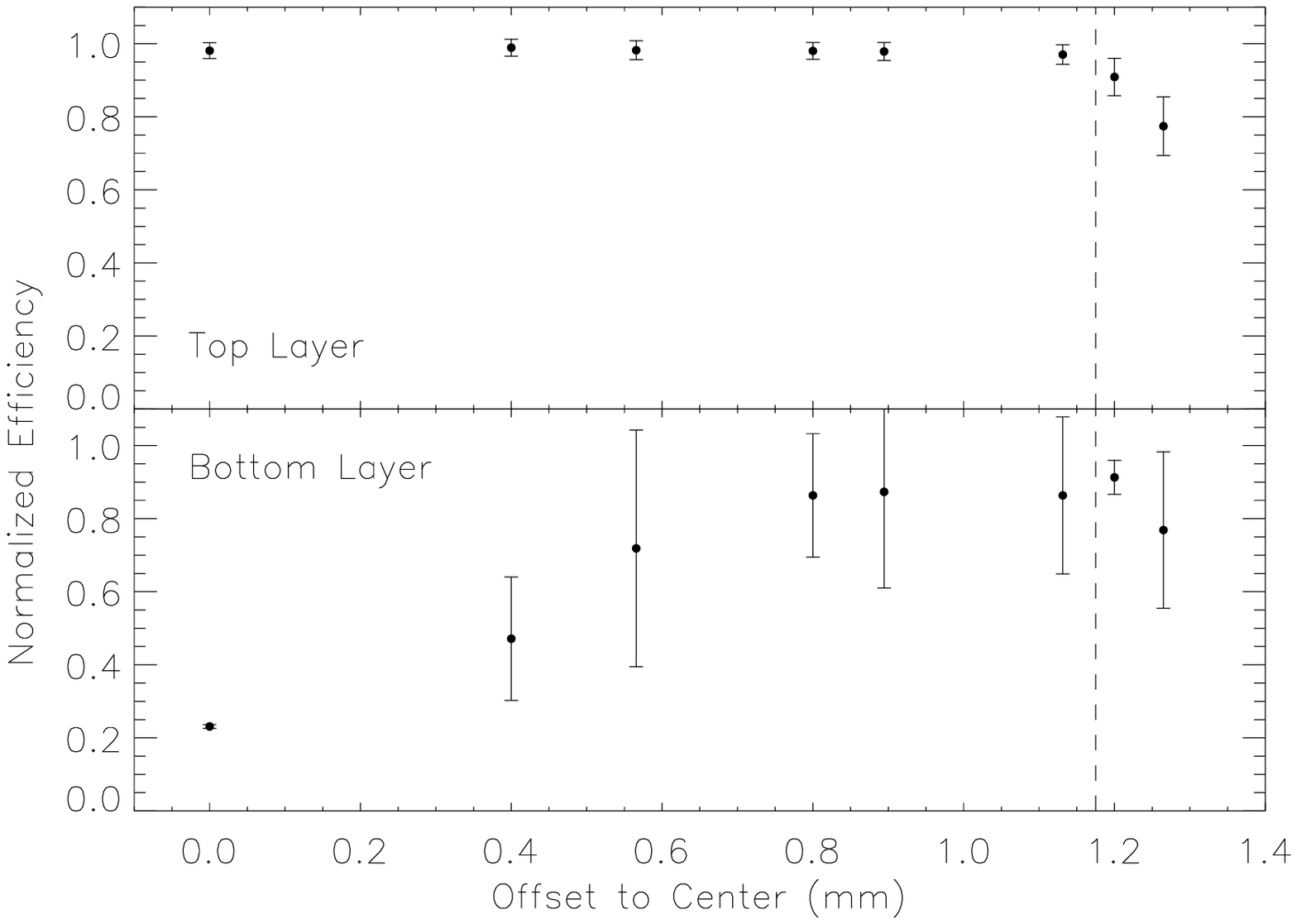}\\
\includegraphics[width=0.49\textwidth]{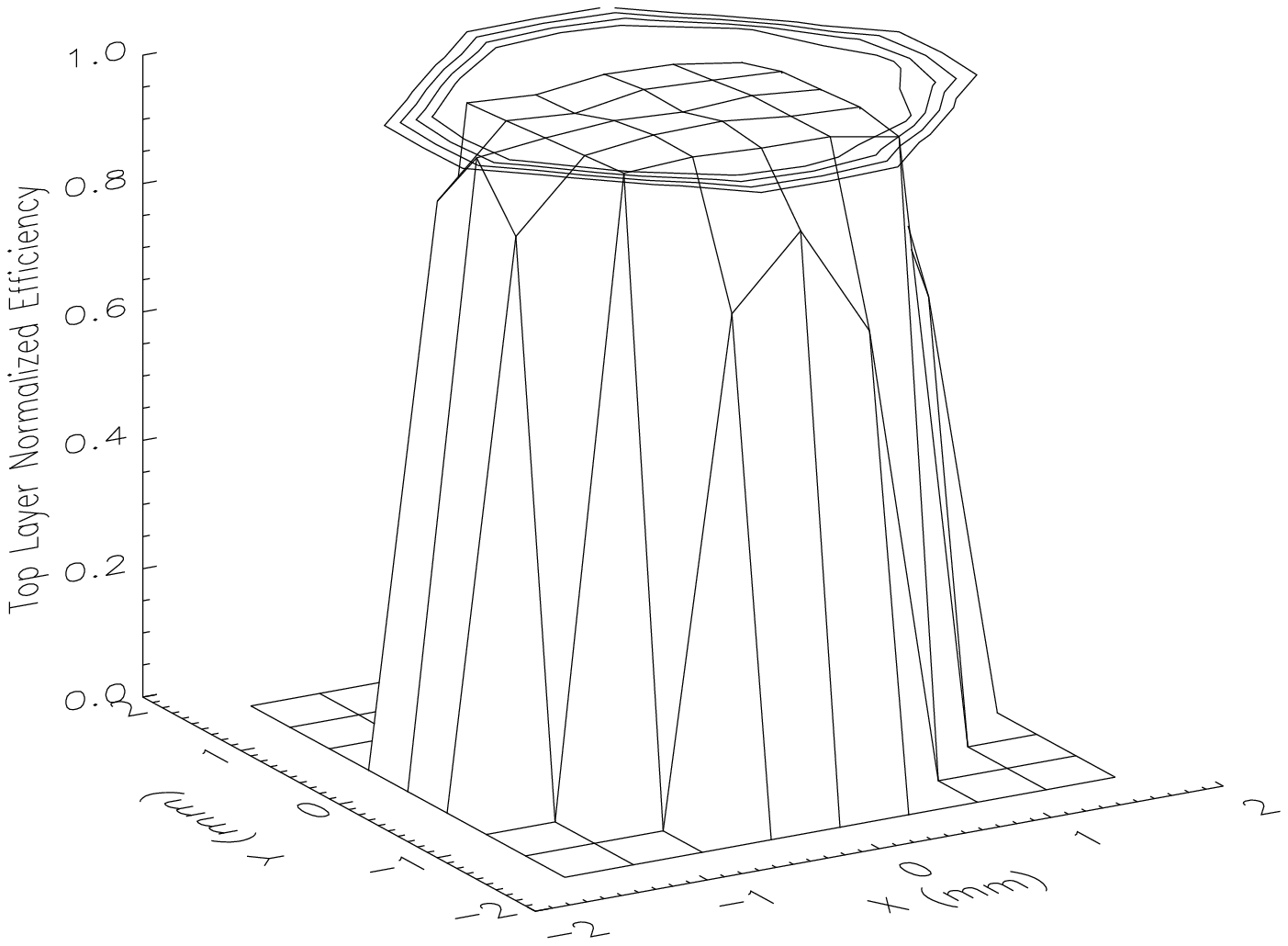}
\includegraphics[width=0.49\textwidth]{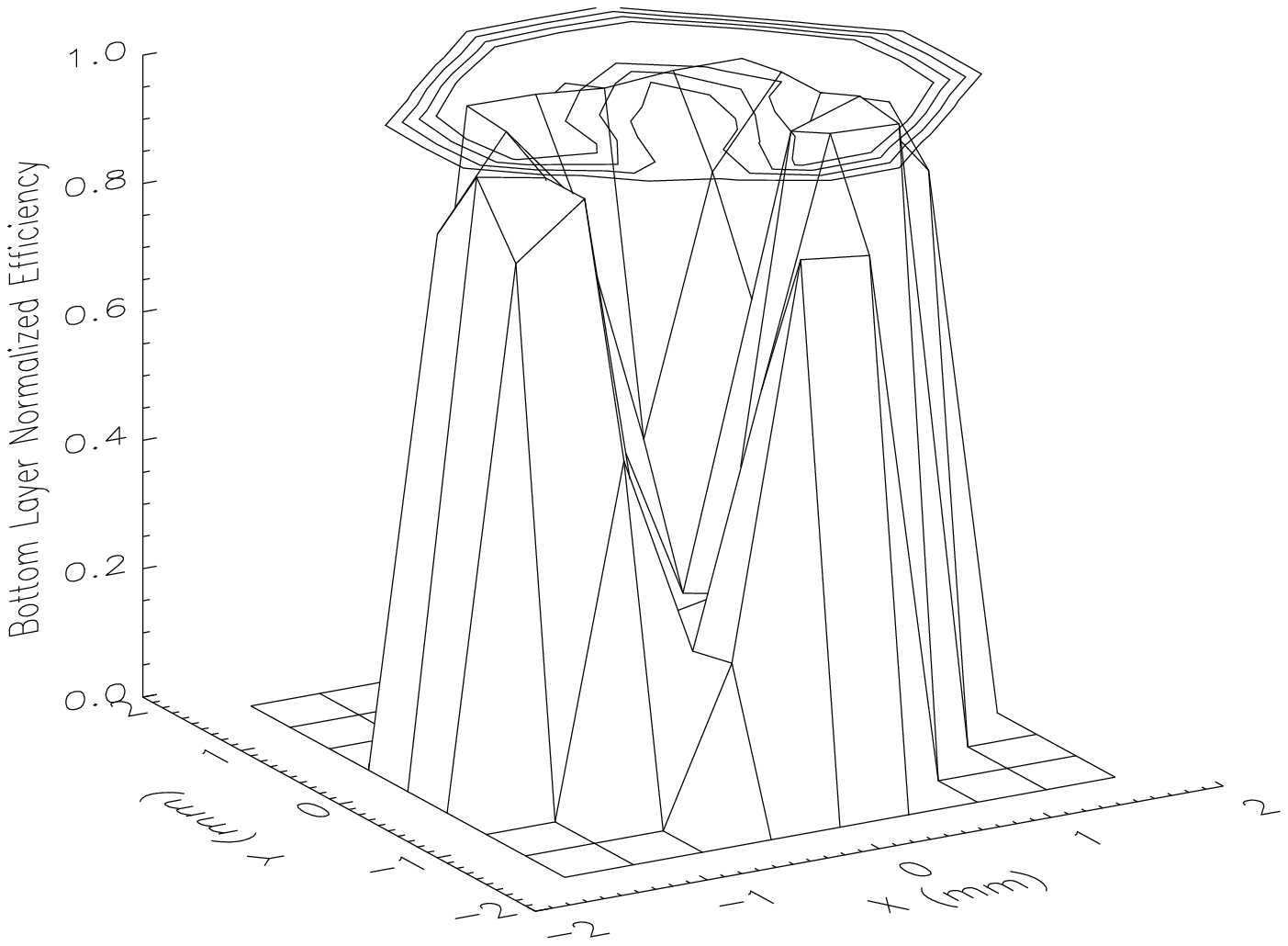}
\caption{Uniformity of efficiency at different points across the
sensitive area. Top Left: crosses are positions from which spectra are
taken from the X-ray tube. The solid circle is the sensitive area of
the detector and the dashed line encircles full response points. Top
Right: local efficiency vs. offset to the center. The dashed line
indicates the full response radius. Bottom Left: local efficiency at
different spots for the top layer. Bottom Right: local efficiency at
different spots for the bottom layer. 
\label{fig:uni}}
\end{figure}

To test the uniformity at different spots on the silicon detector, we
placed a stainless steel mask with a 0.33~mm diameter pinhole in front
of the detector and illuminated it using the X-ray tube to create a
collimated beam.  The mask was mounted on a 2-dimensional linear stage
enabling us to scan the pinhole across the silicon surface.  We
selected 37 points evenly spaced by 0.4~mm to cover the sensitive area
of the detector (see the top left panel of Fig.~\ref{fig:uni}). The
solid circle indicates the sensitive area and the dashed circle
encircles points collimating X-rays all inside the sensitive area, e.g.
the full response region.

The efficiency is measured as the total counting rate in 3-30 keV band.
An alternative measurement using only the characteristic emission lines
from the tube produced the same result. The 2-dimensional efficiencies
for both layers are plotted on the two bottom panels of
Fig.~\ref{fig:uni}. The top player is highly uniform.  The bottom
layer has a central area with low efficiency, which is caused by the Au
high voltage strip which was incorrectly placed into the sensitive area
in the assembly.  The efficiency versus offset to the center is plotted
in the top right panel of Fig.~\ref{fig:uni}.  A 2\% systematic error
is added to account for the instability of the X-ray tube flux.

We also measured the uniformity of charge collection and energy
resolution.  In top panel of Fig.~\ref{fig:unie}, the Mo K$\alpha$ peak
channels are measured at different spots on the detector for both top
and bottom layer.  The shift of the peak channel reflects the charge
collection efficiency.  For both layers, it is shown that the shift is
within 5 channels, corresponding to 70~eV for the 17.44 keV peak.  The
variation is less than 0.5\% and is much smaller than the FWHM.  The
variation of the resolution is within 20~eV except one point in the
bottom layer which is nearly 40~eV.  These changes are very small
indicating excellent uniformity of charge collection and energy
resolution.

\section{Discussion and conclusion}

We have demonstrated a large-format silicon PIN photodiode which
presents good performance in sensitivity, spectroscopy and timing.  The
detector has high quantum efficiency from 1-30 keV, fine spectral
resolution of 220~eV at $-$35 $^\circ$C and 760~eV at 25 $^\circ$C, and
good time resolution of several tens of microseconds or better. 

The intended purpose of the double layer structure was to test
background rejection by coincidence in both layers.  Spectra taken
using a $^{60}$Co source to simulate the radiation environment on-orbit
show that only a small fraction of the background events deposit energy
on both layers simultaneously.  However, the Au foil incorrectly
inserted between the two detector layers, which gave bad uniformity for
the bottom layer, decreased the efficiency for capturing coincident
events.  The on-orbit background counting rate for Si detectors and
means for its reduction are an important topic for future study.

The top layer of the detector shows a excellent uniformity of
efficiency, charge collection and spectral resolution.  This suggests
that we would obtain similar excellent uniformity for Si PIN photodiode
detectors consisting of arrays of small pixels and operated as a
position sensitive detector.

\section{Acknowledgements} 

HF and PK thank The University of Iowa for support.

\clearpage

\begin{table}
\centerline{Table 1: Detected line energies and widths for top and bottom layers.}
\begin{tabular}{rcccccccc}
\noalign{\smallskip} \hline \hline
             & Al K$\alpha\beta$ 
	            & Si K$\alpha$ & Mo L$\alpha\beta$ & Mn K$\alpha$ &  Mn K$\beta$ & Cu K$\alpha$ \\ \hline
Energy (keV) &  1.5 & 1.74 & 2.23 & 5.90 & 6.49 &  8.04 \\
Top ({\tiny FWHM}, eV) &...&...&...& 282  &...&...\\
Bottom ({\tiny FWHM}, eV)& ...& ...  &...&...&...&...\\
\noalign{\smallskip} \hline
& Au L$\alpha$ & Au L$\beta$ & Au L$\gamma$ & Mo K$\alpha$ & Mo K$\beta$ & Ag K$\alpha$ & Ag K$\beta$\\
energy (keV) & 9.71 & 11.48 & 13.38 & 17.44 & 19.61 & 22.10 & 24.94\\
top ({\tiny FWHM}, eV) &...&...&...& 359 &...& 424 &...\\
bottom ({\tiny FWHM}, eV)&...&...&...& 376 &...& 447 &...\\

\noalign{\smallskip} \hline
\end{tabular}
\end{table}

\end{document}